\documentclass[a4paper,10pt,onecolumn,twoside]{cpc-hepnp}
\usepackage{upgreek,fancyhdr}
\usepackage{multicol}
\usepackage{multirow}
\usepackage{graphicx}
\usepackage{booktabs}
\usepackage{amssymb,bm,mathrsfs,bbm,amscd}
\usepackage[tbtags]{amsmath}
\usepackage{lastpage}
\usepackage[english]{babel}
\usepackage{epsfig}
\usepackage{graphicx}
\usepackage{epstopdf}
\usepackage{graphbox}
\usepackage{xcolor}
\usepackage{enumitem} 
\setlist[itemize]{itemsep=0.5em, parsep=0.5em}  

\begin{document}

\fancyhead[c]{\small Chinese Physics C~~~Vol. xx, No. x (2025) xxxxxx}
\fancyfoot[C]{\small xxxxxx - \thepage}
\footnotetext[0]{Received \today}

\title{Revised classification of the CHIME fast radio bursts with machine learning\thanks{Supported by the National Natural Science Fund of China under grant Nos. 12275034 and 12347101, and the Natural Science Fund of Chongqing under grant No. CSTB2022NSCQ-MSX0357.}}

\author{Liang Liu$^{1}$
\quad Hai-Nan Lin$^{2;1)}$\email{linhn@cqu.edu.cn}
\quad Li Tang$^{1}$}

\maketitle
\hspace{1cm}

\address{$^1$ School of Physics and Electronic Information, Mianyang Teachers' College, Mianyang 621000, China\\
$^2$ Department of Physics, Chongqing University, Chongqing 401331, China}


\begin{abstract}
  Fast radio bursts (FRBs) are short-duration and energetic radio transients of unknown origin. Observationally, they are commonly categorized into repeaters and non-repeaters. However, this binary classification may be influenced by observational limitations such as sensitivity and time coverage of telescopes. In this work, we employ unsupervised machine learning techniques to re-examine the CHIME/FRB catalog, with the goal of identifying intrinsic groupings in the FRB population without relying on preassigned labels. Using t-distributed stochastic neighbor embedding (t-SNE) for dimensionality reduction and hierarchical density-based spatial clustering of applications with noise (HDBSCAN) for clustering, we find that the FRB sample separates naturally into two major clusters. One cluster contains nearly all known repeaters but is contaminated by some apparently non-repeaters, while the other cluster is dominated by non-repeaters. This suggests that certain FRBs previously labeled as non-repeaters may share intrinsic similarities with repeaters. The mutual information analysis reveals that rest-frame frequency width and peak frequency are the most informative features governing the clustering structure. Even when reducing the input space to just these two features, the classification remains robust. 
\end{abstract}



\footnotetext[0]{\hspace*{-3mm}\raisebox{0.3ex}{$\scriptstyle\copyright$}2019
Chinese Physical Society and the Institute of High Energy Physics
of the Chinese Academy of Sciences and the Institute
of Modern Physics of the Chinese Academy of Sciences and IOP Publishing Ltd}%

\section{Introduction}\label{sec:introduction}

Fast radio bursts (FRBs) are millisecond-duration bursts of radio waves, first identified in 2007 from the archival data of the Parkes radio telescope \cite{Lorimer:2007qn}. To date, more than 1000 FRB sources have been detected by telescopes worldwide \cite{Xu:2023did}, with publicly available data repositories providing extensive catalogs\footnote[2]{www.wis-tns.org}. These enigmatic transients have emerged as powerful tools for probing astrophysical and cosmological phenomena, yet their origins remain unresolved. The majority of FRBs exhibit high dispersion measures (DMs) that exceed the expected contribution from the Milky Way, indicating extragalactic or even cosmological distances \cite{Keane:2016yyk,Chatterjee:2017dqg}. Their observed properties, along with polarization and spectral measurements, offer crucial insights into the physical mechanisms underlying their emission. In particular, the discovery of repeating FRBs, most notably FRB 121102 \cite{Scholz:2016rpt,Spitler:2016dmz}, has provided key constraints on progenitor models, suggesting at least a subset of FRBs arise from non-cataclysmic sources. However, despite the numerous theoretical models proposed to explain their emission mechanisms, a unifying theory capable of accounting for the full diversity of FRB phenomena remains elusive \cite{Platts:2018hiy,Zhang:2020qgp}.

FRBs have traditionally been categorized into two primary categories \---- repeating and non-repeating \---- based on the number of detected bursts. Distinct observational characteristics between these groups suggest the possibility of heterogeneous origins. Among repeating FRBs, only a small fraction exhibit high burst rates \cite{Li:2021hpl,Xu:2021qdn,Niu:2021bnl}, while the majority display limited activity. It remains plausible that some FRBs classified as non-repeaters are, in fact, repeaters whose bursts have not been observed yet due to insufficient monitoring or sensitivity limitation, raising concerns about potential misclassification. Such uncertainties complicate efforts to constrain progenitor models and decipher the underlying emission mechanisms. Furthermore, emerging evidence suggests that FRBs may not be strictly dichotomous, with additional subpopulations potentially existing \cite{Chen:2021jpq,Yang:2023dcf,Sun:2024huw,Luo:2022smj,Zhu-Ge:2022nkz}. A more refined classification scheme is therefore essential for resolving their origins, improving theoretical modeling, and advancing our understanding of their astrophysical nature.

Although previous studies have reported significant differences in the parameter distributions of repeating and non-repeating FRBs, conventional comparative analyses have largely been confined to individual parameters or low-dimensional projections \cite{CHIMEFRB:2019pgo,Fonseca:2020cdd,CHIMEFRB:2021srp,Li:2021hpl,Zhong:2022uvu}. Such approaches, however, are inherently limited in their ability to capture the complex, potentially high-dimensional correlations that may underpin the observed diversity of FRB properties. To address these limitations, recent efforts have increasingly turned to machine learning \cite{Chen:2021jpq,Yang:2023dcf,Sun:2024huw,Luo:2022smj,Zhu-Ge:2022nkz,Raquel:2023tgi,Garcia:2024ckc,Qiang:2024lhu} \---- a branch of artificial intelligence well-suited to extracting hidden patterns from large, multi-dimensional datasets. Machine learning techniques have been widely applied in cosmological and astronomical research, and can be broadly classified into two categories: supervised and unsupervised learning, distinguished by whether or not the data are labeled. In supervised learning, models are trained on labelled datasets to infer explicit mappings between inputs and target outputs, allowing prior knowledge to guide predictive accuracy. In contrast, unsupervised learning operates without labeled data, seeking to uncover intrinsic structures, patterns or clusters within the data.

Machine learning techniques have demonstrated considerable promise in the classification of FRBs, although the resulting classifications can vary depending on the choice of methods and input parameters \cite{Chen:2021jpq, Yang:2023dcf, Luo:2022smj, Zhu-Ge:2022nkz, Sun:2024huw,Luo:2025}. Employing a range of supervised learning algorithms, Luo et al. \cite{Luo:2022smj} analyzed FRBs from the first Canadian Hydrogen Intensity Mapping Experiment Fast Radio Burst (CHIME/FRB) catalog, dividing the data into training and testing sets to evaluate model performance. Their approach enabled the identification of several candidate repeaters within the non-repeating FRB population. In contrast, Chen et al. \cite{Chen:2021jpq} applied an unsupervised learning technique \---- Uniform Manifold Approximation and Projection (UMAP) \---- to the same dataset, achieving a repeater completeness of 95$\%$ and identifying 188 candidate repeaters from 474 non-repeating sources. Motivated by concerns that the sensitivity of Chen et al.'s method may be affected by the choice of the hyperparameter \texttt{n\_neighbors}, Zhu-Ge et al. \cite{Zhu-Ge:2022nkz} revisited the CHIME/FRB catalog using UMAP alongside two additional dimensionality-reduction techniques: principal component analysis (PCA) and t-distributed stochastic neighbour embedding (t-SNE). Their analysis revealed the presence of not only the two conventional classes of repeating and non-repeating FRBs, but also three additional subgroups, suggesting a more complex underlying classification structure. By applying an unsupervised decision tree algorithm to the first CHIME/FRB catalog, Luo et al.\cite{Luo:2025} observed a reduced distinction between repeaters and non-repeaters upon inclusion of the CHIME/FRB 2023 catalog.

Previous classification efforts have typically treated each burst \---- whether a sub-burst from a non-repeating FRB or an individual burst from a repeating source \---- as an independent event. However, bursts originating from the same FRB source are likely to share intrinsic physical correlations. Ignoring these correlations and treating bursts as independent during model training risks introducing redundant information, which may bias the learning process towards overfitting and hinder the model's ability to capture the global properties of FRB sources. To mitigate these issues and improve the robustness of the classification, we employ unsupervised machine learning methods, including dimensionality reduction and clustering, to analyze the CHIME/FRB catalogs, retaining only a single representative burst from each source. Specifically speaking, we only preserve the first detected burst from each repeating source, and the first sub-burst from each non-repeating source, while neglecting all subsequent (sub-)bursts from the same source. Our approach to repeating FRBs is aligned with that proposed by Zhong et al. \cite{Zhong:2022uvu}.

This paper is structured as follows. In Section {\ref{sec:data and method}}, we describe the datasets used in this study and outline the unsupervised machine learning techniques adopted for dimensionality reduction and clustering. In section {\ref{sec:results}}, we present the main results of our analysis. Finally, discussions and conclusions are provided in Section {\ref{sec:conclusions}}.

\section{Data and methodology}\label{sec:data and method}

\subsection{Data selection}\label{subsec: data}

The dataset utilized in this study is primarily derived from the first CHIME/FRB catalog \cite{CHIMEFRB:2021srp} and the CHIME/FRB 2023 catalog \cite{CHIMEFRB:2023myn}. The first CHIME/FRB catalog contains 536 bursts detected between 400 and 800 MHz during the period from July 25, 2018 to July 1, 2019. Of those, 474 events originate from apparently non-repeating sources and 62 events are associated with 18 known repeating sources. Among the non-repeating bursts, 64 exhibit sub-burst structures. Six FRBs exhibiting zero values in both fluence and flux \---- FRB20190307A, FRB20190307B, FRB20190329B, FRB20190329C, FRB20190531A, and FRB20190531B \---- are removed from the sample to ensure robustness in the subsequent analysis. Therefore, the first CHIME/FRB catalog contains $474+18-6=486$ available independent FRB sources. The CHIME/FRB 2023 repeater catalog comprises 25 repeating FRB sources detected between September 30, 2019 and May 1, 2021, including 6 sources that were previously classified as non-repeaters in the first CHIME/FRB catalog. To mitigate bias from intrinsic burst correlations within sources, we retain only the chronologically earliest burst per repeating source and the first sub-burst per non-repeating source, while excluding all subsequent (sub-)bursts from the same source. This ensures that exactly one event per source is utilized. The resultant dataset contains 505 independent bursts, among which 43 originate from repeating sources and 462 from non-repeating sources.

\begin{figure}[ht!]
\centering
\includegraphics[width=0.32\textwidth]{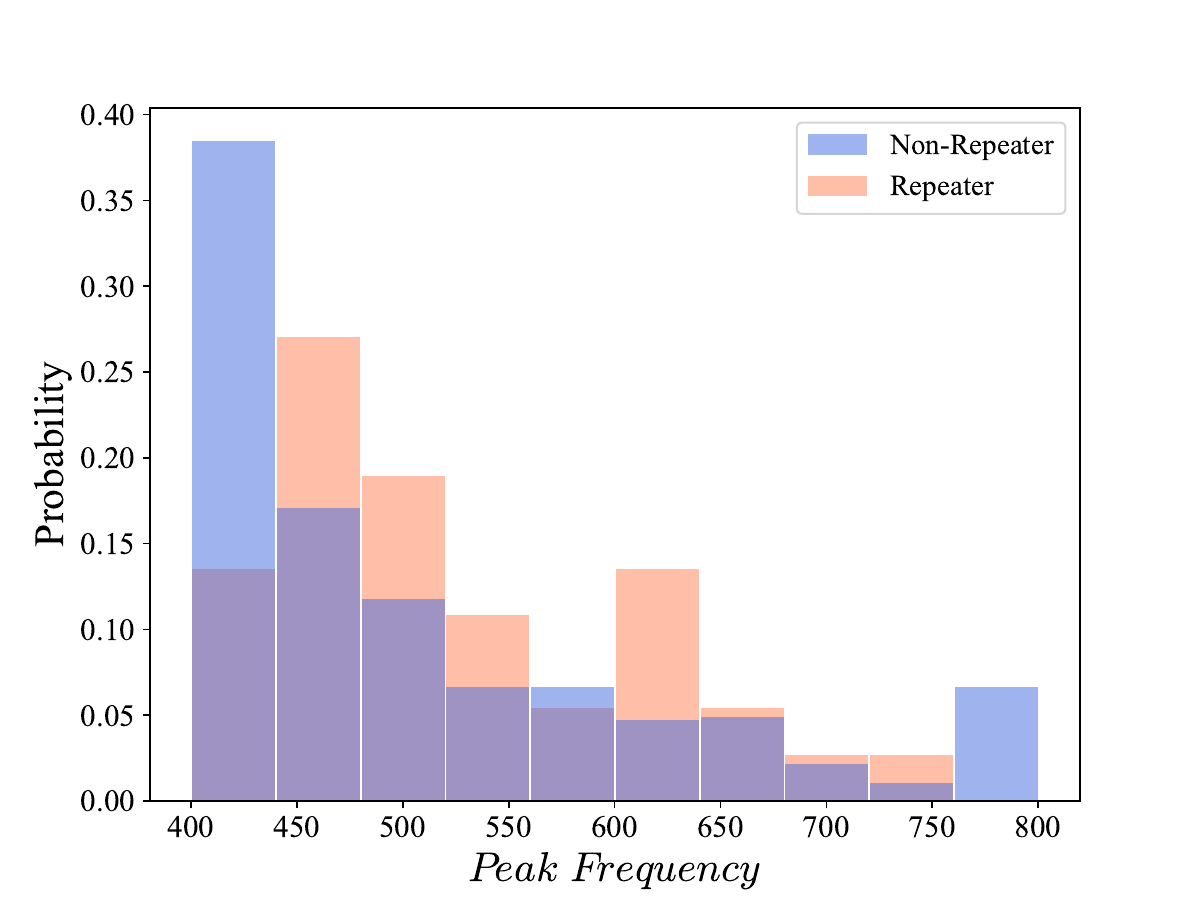}
\includegraphics[width=0.32\textwidth]{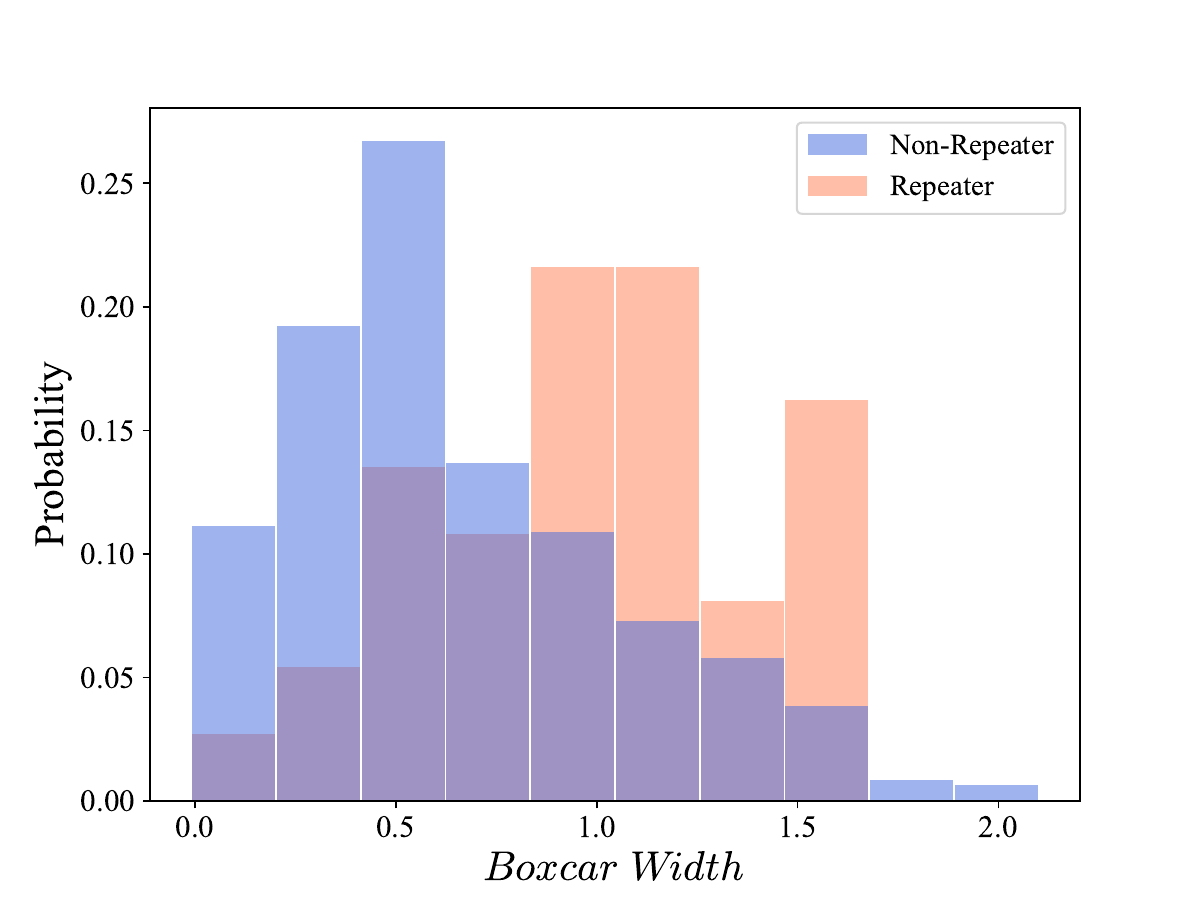}
\includegraphics[width=0.32\textwidth]{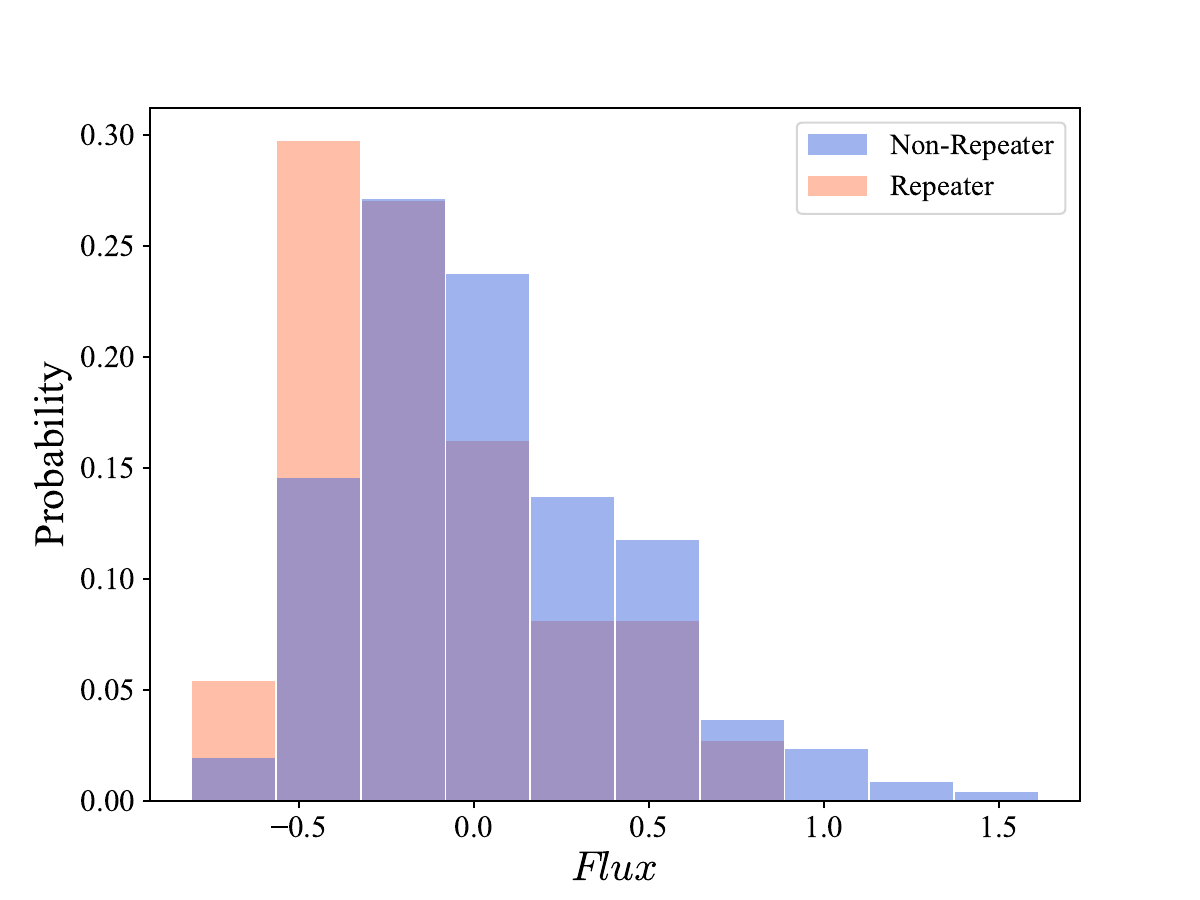}
\includegraphics[width=0.32\textwidth]{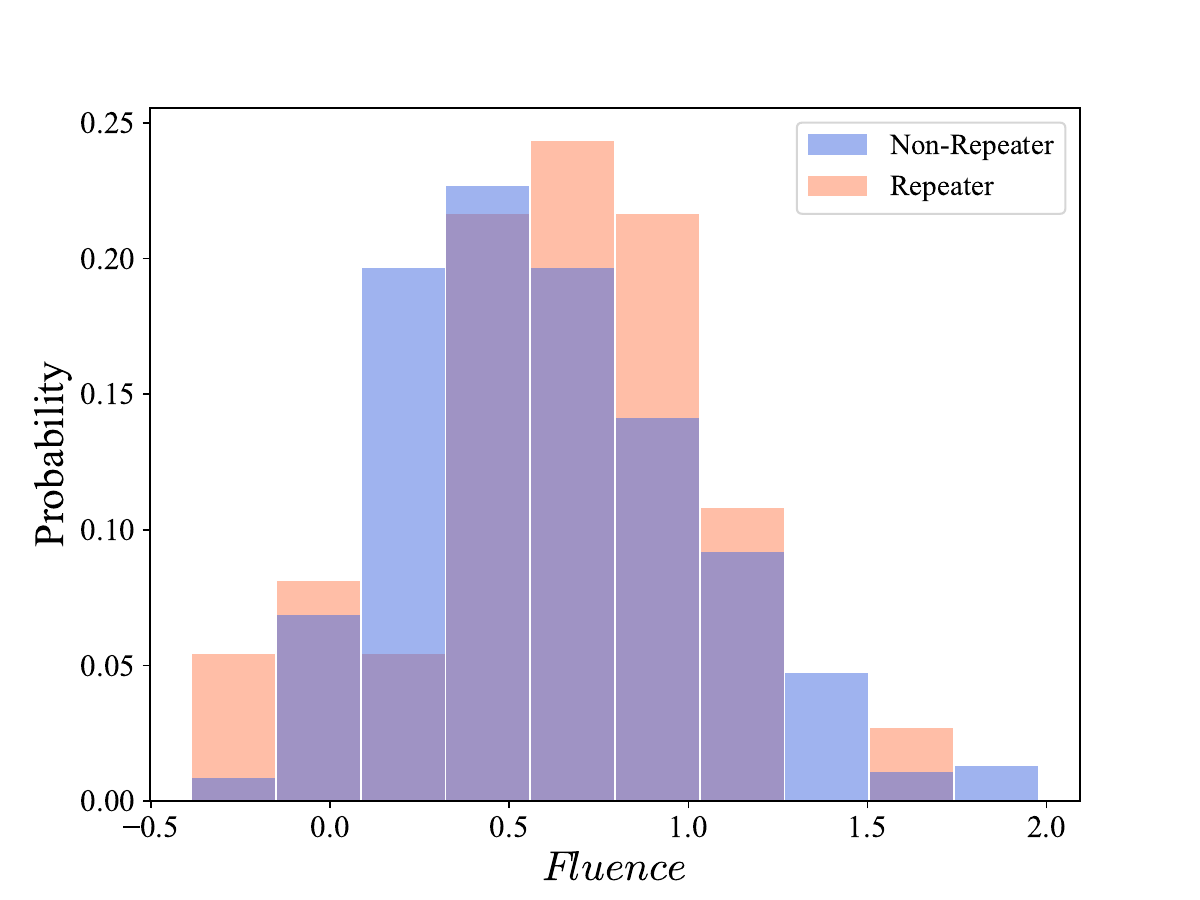}
\includegraphics[width=0.32\textwidth]{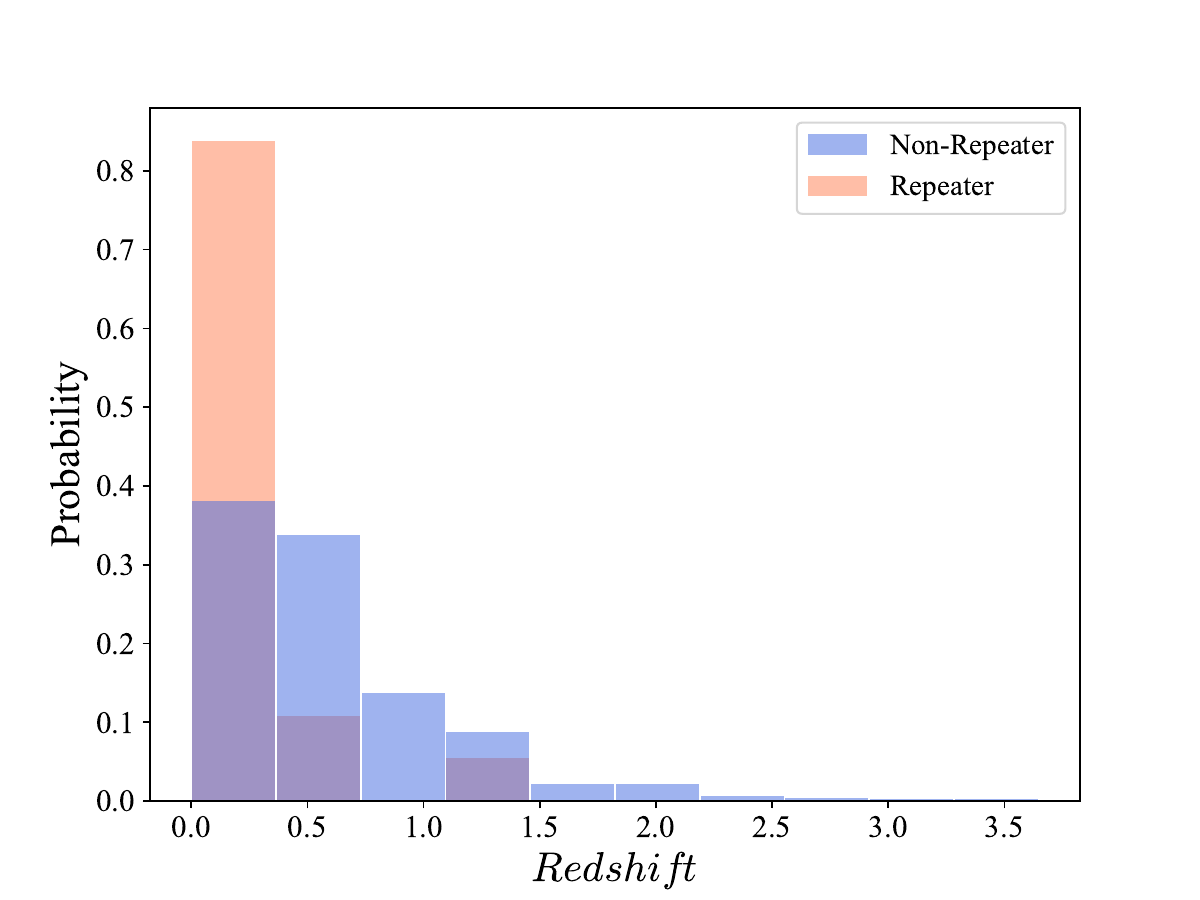}
\includegraphics[width=0.32\textwidth]{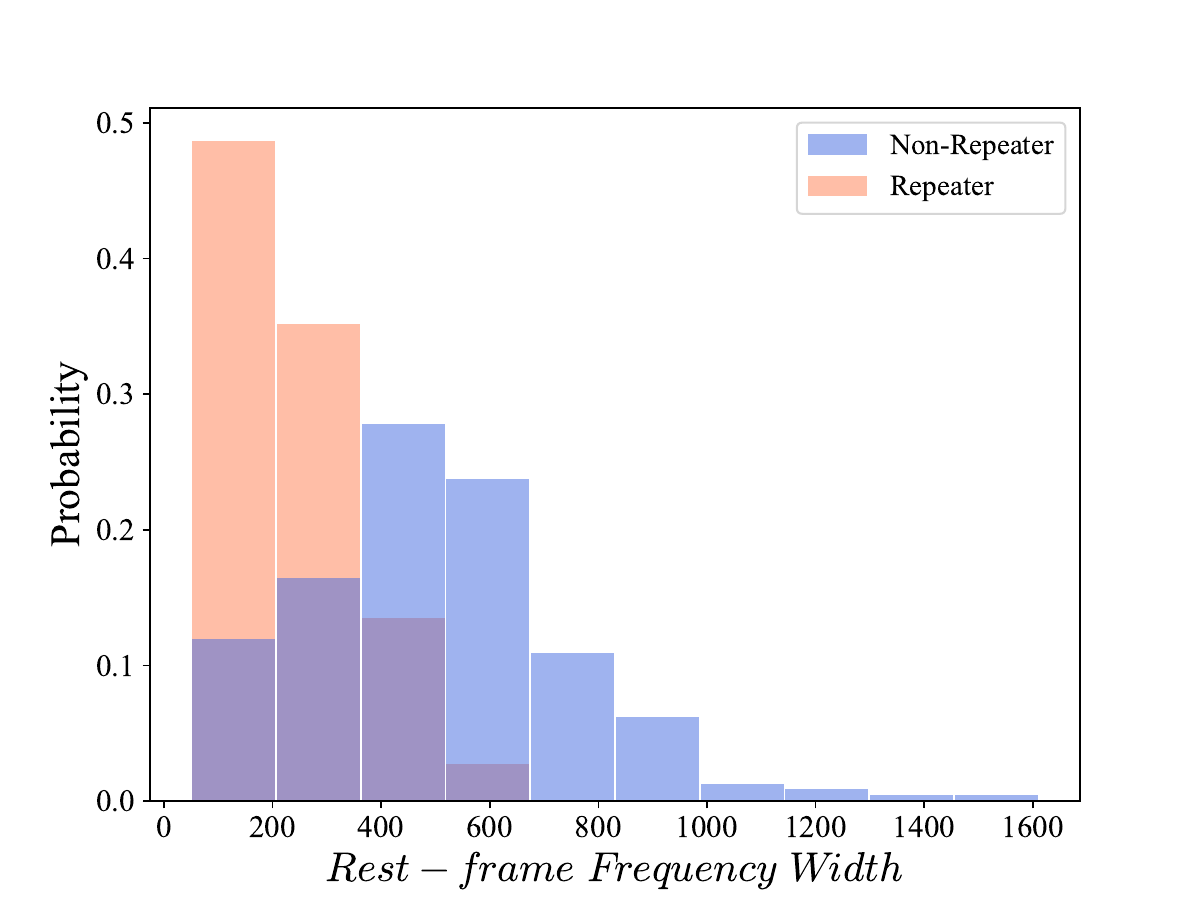}
\includegraphics[width=0.32\textwidth]{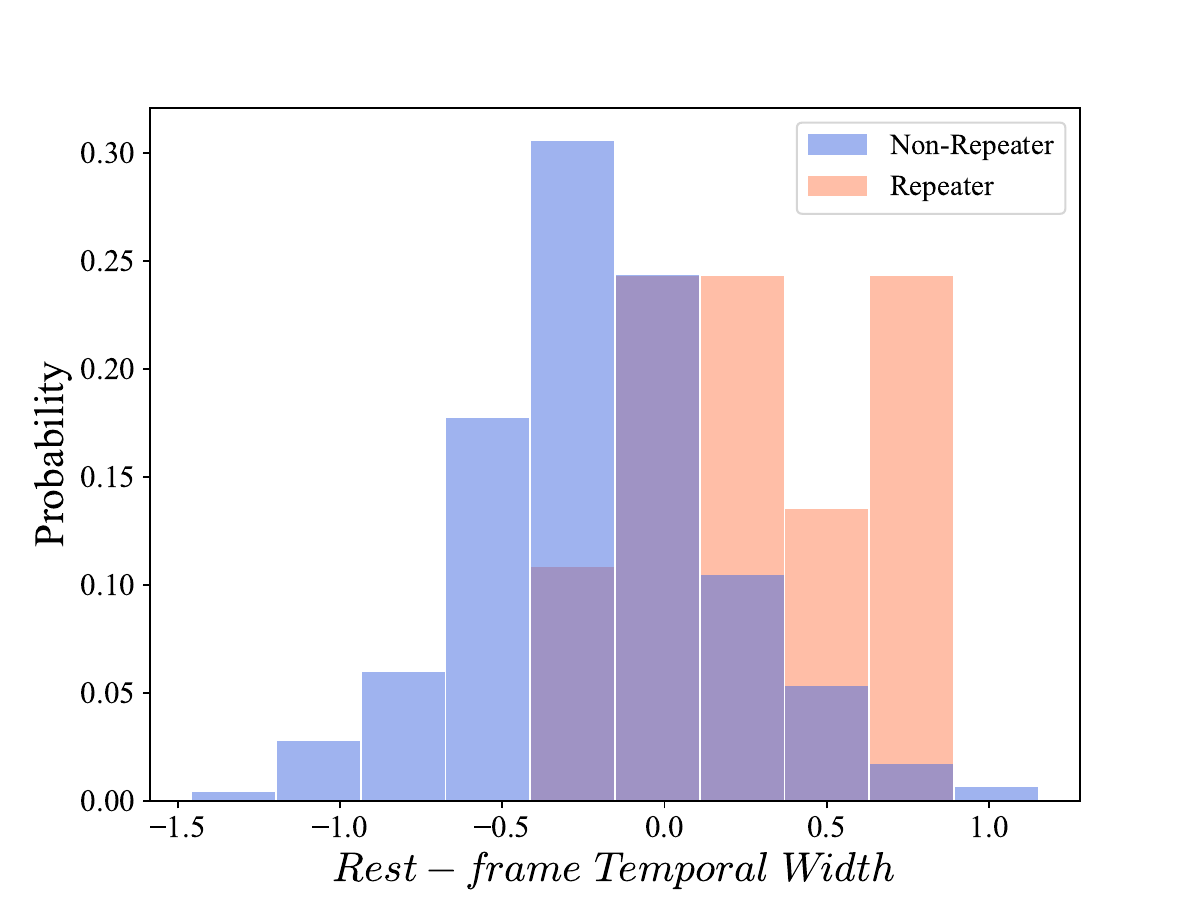}
\includegraphics[width=0.32\textwidth]{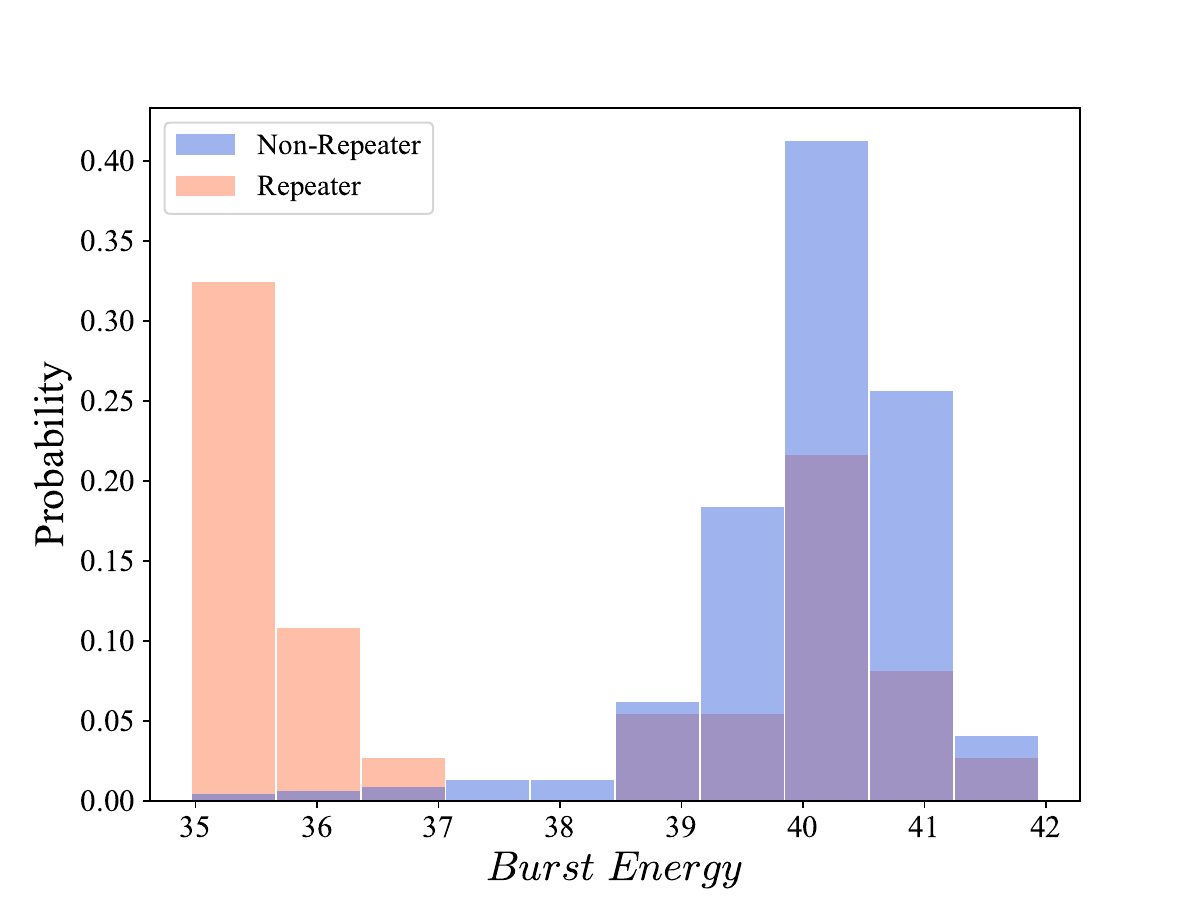}
\includegraphics[width=0.32\textwidth]{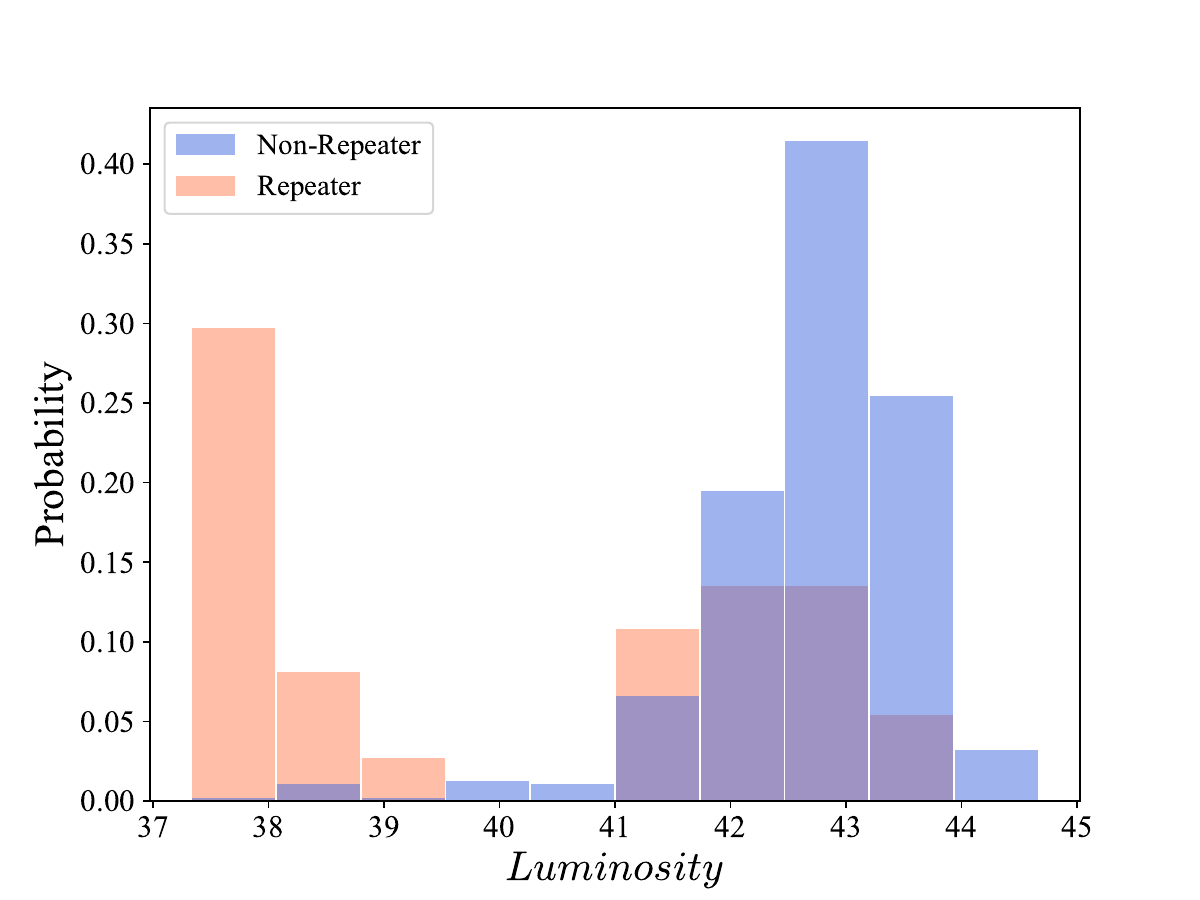}
\includegraphics[width=0.32\textwidth]{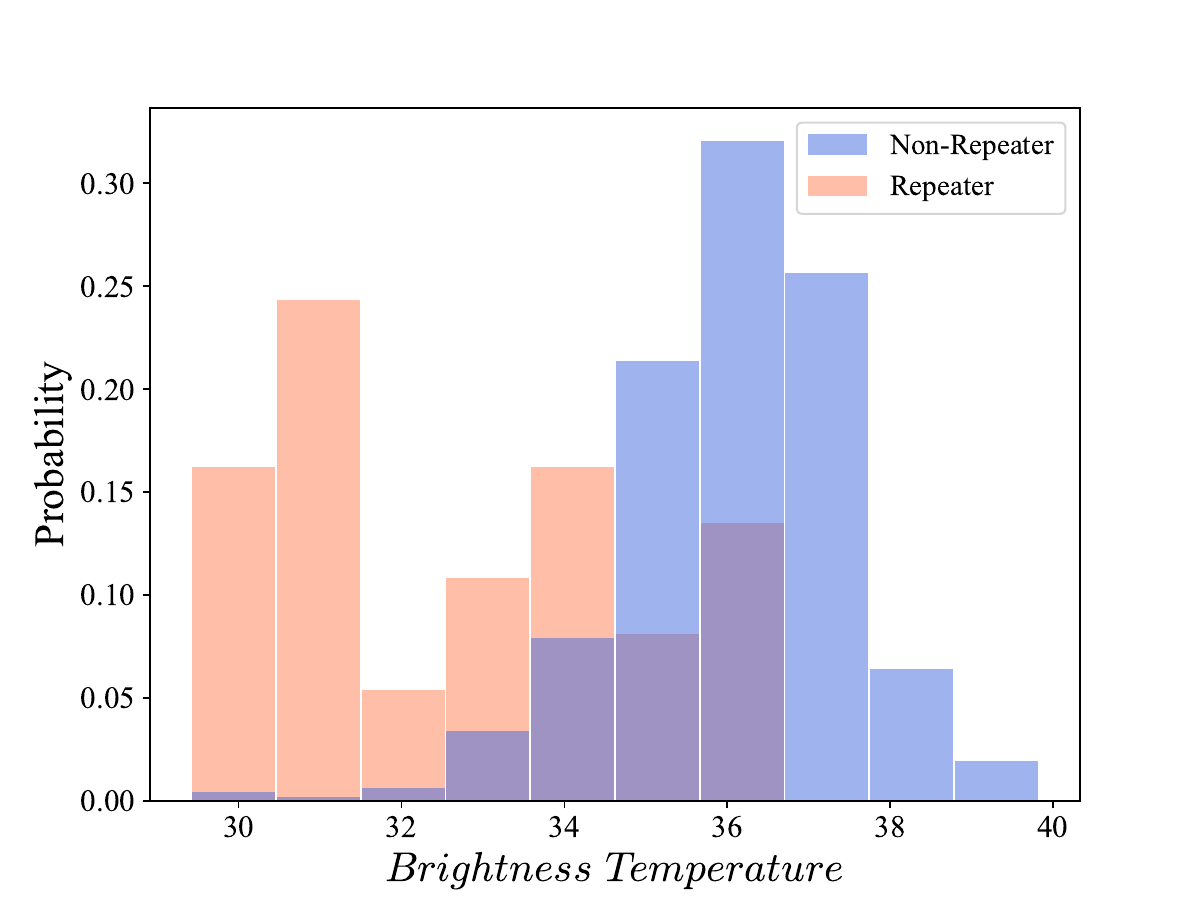}
\caption{The histograms of ten features that are used in the classification. The vertical axis is normalized such that the total probability is unity.}
\label{fig:dist}
\end{figure}

In our unsupervised machine learning framework, ten features are extracted either directly from the observed properties reported in the CHIME/FRB catalog or derived through standard calculations. These features include: peak frequency, flux density, fluence, boxcar width, redshift, rest-frame frequency width, rest-frame temporal width, burst energy, luminosity, and brightness temperature. The distributions of these features are shown in Fig.\ref{fig:dist}. Detailed definitions and descriptions of each feature are provided below.

\begin{itemize}

\item{Peak frequency $\nu_p$ (MHz). The peak frequency is defined as the frequency at which the sub-burst reaches its maximum flux density.}

\item{Flux $S_{\nu}$(Jy). The flux density reported in the catalog corresponds to the peak flux of the band-averaged profile and represents a lower-limit estimate. Logarithmic values are used in the analysis.}

\item{Fluence $F_{\nu}$(Jy ms). The fluence is defined as the time-integrated flux density of the burst, as provided in the CHIME/FRB catalog. Logarithmic values are adopted.}

\item{Boxcar width $\Delta t_{\rm BC}$ (ms). The burst duration is characterized by the width of the boxcar function after convolution, as defined in the catalog. Logarithmic values are used in the analysis.}

\item{Redshift $z$. Most FRBs in our sample lack direct redshift measurements, with the exception of two sources \---- FRB20121102A ($z=0.19273$) \cite{Tendulkar:2017vuq} and FRB20180916B ($z=0.0337$) \cite{Marcote:2020ljw} \---- for which spectroscopic redshifts are available via host galaxy associations. For the remaining FRBs, redshifts are estimated using the empirical relation between dispersion measure (DM) and redshift, which will be described in details below.}

\item{Rest-frame frequency width $\Delta \nu$ (MHz). The rest-frame frequency width characterizes the intrinsic spectral extent of each burst and is calculated as the redshift-corrected difference between the maximum and minimum observed frequencies,
\begin{equation}
    \Delta\nu=(\nu_{\rm max}-\nu_{\rm min})(1+z).
\end{equation}
}

\item{Rest-frame temporal width $\Delta t_{r\rm }$ (ms). The rest-frame temporal width of each sub-burst is determined using fitburst $\Delta t$ \cite{CHIMEFRB:2021srp}, corrected for cosmological time dilation,
\begin{equation}
    \Delta t_{\rm r}=\frac{\Delta t}{1+z}.
\end{equation}
Logarithmic values of these rest-frame temporal widths are used in the analysis.}

\item{Burst energy $E$ (erg). The burst energy of each FRB is estimated as
\begin{equation}
    E=\frac{4\pi D_{L}^2}{1+z}F_{\nu}\nu_p,
\end{equation}
where $F_{\nu}$ is the specific fluence, $D_L$ is the luminosity distance, and $\nu_p$ is the observed peak frequency of the burst. Logarithmic values of the inferred energies are adopted for subsequent analysis.}

\item{Luminosity $L$ (erg s$^{-1}$ ). The luminosity of FRBs is estimated as
\begin{equation}
    L=4\pi D_L^2 S_{\nu}\nu_p,
\end{equation}
where $S_{\nu}$ denotes the specific peak flux density. Logarithmic values of the derived luminosities are adopted in the analysis.}

\item{Brightness temperature $T_B$ (K). The brightness of a source is characterized by its brightness temperature, defined as the temperature of a blackbody emitting the same specific intensity. Accounting for cosmological effects, $T_B$ is calculated as \cite{stac3206}
\begin{equation}
    T_B=\frac{S_{\nu} D_A^2}{2\pi k_B(\nu_p \Delta t)^2}(1+z)^3,
\end{equation}
where $k_B$ is the Boltzmann constant, $\Delta t$ is the observed duration, and $D_A$ is the angular diameter distance. Logarithmic values are used in the analysis.}

\end{itemize}

In order to infer the redshift of unlocalized FRBs from the DM, we decompose the observed DM of an extragalactic FRB into four components as usual \cite{Macquart:2020lln,Deng:2013aga,Gao:2014iva},
\begin{equation}\label{eq:DM}
    {\rm {DM}}={\rm DM_{MW}}+{\rm DM_{halo}}+{\rm DM_{IGM}}+\frac{{\rm DM_{host}}}{1+z},
\end{equation}
where the four terms on the right-hand-side denote the contributions from the Milky Way interstellar medium, the galactic halo, the intergalactic medium, and the host galaxy, respectively. The Milky Way contribution, $\rm DM_{MW}$, is estimated using the NE2001 electron density model based on pulsar observations \cite{Cordes:2002wz}. Following the work of Zhu-Ge et al. \cite{Zhu-Ge:2022nkz}, the galactic halo and host galaxy terms are fixed to $\rm DM_{halo}$ = 30 pc cm$^{-3}$ and  $\rm DM_{host}$  = 70 pc cm$^{-3}$, respectively. The contribution from the intergalactic medium is computed within a flat $\Lambda$CDM cosmology as
\begin{equation}\label{eq:DM_IGM}
    {\rm DM_{IGM}}(z)=\frac{21cH_0\Omega_bf_{\rm IGM}}{64\pi Gm_p}\int_0^z\frac{1+z}{\sqrt{\Omega_m(1+z)^3+\Omega_\Lambda}}dz,
\end{equation}
where $c$ is the speed of light, $G$ is Newton's gravitational constant, $m_p$ is the proton mass, and $f_{\rm IGM}$ = 0.83 is the baryon fraction of the IGM \cite{Fukugita:1997bi}. The cosmological parameters are set to the Planck 2018 values: $H_0=67.4~{\rm km~s^{-1}~Mpc^{-1}}$, $\Omega_m=0.315$, $\Omega_\Lambda=0.685$ and $\Omega_{b}=0.0493$ \cite{Planck:2018vyg}. Redshifts are then inferred by inverting the relation above given the observed DM. For sources located near the Milky Way, the resulting redshift can approach zero or become formally negative. In such cases, a minimum redshift cutoff of $z = 0.002248$ \---- corresponding to a luminosity distance of 10 Mpc \---- is adopted.

\subsection{Dimensionality reduction}\label{subsec:dim}

While high-precision observations of FRBs offer unprecedented opportunities to probe their enigmatic nature, the inherently high dimensionality of these datasets presents considerable challenges for both classification and physical interpretation. To address this, dimensionality reduction \---- a process that projects data from a high-dimensional space onto a more tractable low-dimensional manifold \---- has become an essential field of machine learning \cite{manifold_Cayton2005}. Dimensionality reduction provides a reasonable visualization in low-dimensional plane to analyze the intrinsic connections of the data. However, conventional linear techniques often fail to preserve the proximity of similar data points in the reduced space, a limitation particularly problematic when the data reside on non-linear, low-dimensional manifolds embedded in higher dimensions. To overcome this, a suite of non-linear dimensionality reduction methods has been developed with an emphasis on preserving local structure. Among these, Stochastic Neighbor Embedding (SNE) has gained prominence for its capacity to generate meaningful low-dimensional representations \cite{sne}. SNE achieves this by converting high-dimensional Euclidean distances into conditional probabilities that quantify pairwise similarity under a Gaussian kernel, and subsequently optimizing a cost function \---- formulated as the Kullback–Leibler divergence between the high- and low-dimensional distributions.

However, like many manifold learning algorithms, SNE suffers from the so-called ``crowding problem". In high-dimensional spaces, a data point may have numerous equidistant neighbours; yet, when projected into a low-dimensional space, there is insufficient room to accommodate all of these neighbours without artificially compressing them or causing overlaps. This spatial constraint distorts the true pairwise relationships, often resulting in a congested central region in the visualization. To mitigate this issue, the t-distributed stochastic neighbour embedding (t-SNE) algorithm replaces the Gaussian kernel used in the low-dimensional space with a student's t-distribution \cite{tsne_visual,tsne_acc}. Owing to its heavier tails, the t-distribution allows for moderate distances in the high-dimensional space to be mapped to larger separations in the low-dimensional embedding. This adjustment effectively alleviates the crowding effect, preserving the relative topology of the data more faithfully.

In this study, we project the input data \---- FRBs characterized by ten features \---- onto a two-dimensional manifold, and accordingly set the \texttt{n\_components} hyperparameter of the t-SNE algorithm to 2. Among the various tunable parameters in t-SNE, the most critical one is the \texttt{perplexity}, which acts as a smooth measure of the effective number of local neighbours considered during the estimation of similarity probabilities in the high-dimensional space. Conceptually, \texttt{perplexity} governs the trade-off between capturing local structure and preserving global relationships: lower values accentuate fine-grained clustering, while higher values incorporate broader context at the potential cost of merging distinct group boundaries. Following the heuristic proposed by Oskolkov \cite{Oskolkov:2019}, \texttt{perplexity} is typically chosen to scale with the square root of the dataset size ($N=505$ in our case). Therefore, we set $\texttt{perplexity}=\sqrt{N}=22$. A summary of all hyperparameter settings employed in the t-SNE dimensionality reduction is summarized in Table \ref{tab:tSNE}.

\begin{table}[htbp]
\centering
\caption{List of t-SNE hyperparameters.}\label{tab:tSNE}
\arrayrulewidth=1.0pt
\renewcommand{\arraystretch}{1.3}
{\begin{tabular}{cc} 
\hline\hline 
Name & Value\\\hline
n$\_$components & 2 \\
perplexity & 22 \\
early$\_$exaggeration & 2 \\
learning$\_$rate & `\ auto' \\
n$\_$iter & 1000 \\
n$\_$iter$\_$without$\_$progress & 300 \\
min$\_$grad$\_$norm & 1e$-$7 \\
metric & `\ euclidean' \\
metric$\_$params & None \\
init & `\ random' \\
verbose & 0 \\
random$\_$state & 22 \\
method & `\ barnes hut' \\
angle & 0.5 \\
n$\_$jobs & None \\
square$\_$distances & `\ deprecated' \\
\hline
\end{tabular}}
\end{table}

\subsection{Clustering method}\label{subsec:cluster}

Following dimensionality reduction, we apply clustering techniques to analyze the resulting low-dimensional representations. Clustering, a core class of unsupervised learning methods, enables the identification of intrinsic structure in unlabeled data by partitioning it into distinct groups, or clusters, such that intra-cluster similarity is maximized while inter-cluster similarity is minimized. Unsupervised clustering automatically classifies the data without the need to preset the number of classes. Among the variety of clustering algorithms, Density-Based Spatial Clustering of Applications with Noise (DBSCAN) offers distinct advantages in handling complex and noisy datasets \cite{dbscan_Ester1996}. DBSCAN defines clusters as contiguous regions of high point density, separated by regions of lower density, and is therefore capable of discovering clusters of arbitrary shape. As a result, DBSCAN is particularly well-suited for analyzing the potentially heterogeneous and irregular structures found in FRB parameter space.

The efficacy of DBSCAN in uncovering meaningful structure within complex datasets hinges on two key hyperparameters: the neighbourhood radius, $\epsilon$, and the minimum number of points, \texttt{minPts}, required to define a dense region. The parameter $\epsilon$ sets the spatial scale at which local density is evaluated, determining the maximum distance within which two points are considered neighbours. The parameter \texttt{minPts} specifies the minimum number of data points, including the point itself, that must reside within the $\epsilon$-neighbourhood for a region to be considered dense. Together, these parameters define the local density criteria that underpin both cluster formation and the identification of noise. Based on these criteria, each data point is classified into one of three categories:

\begin{itemize}
\item{Core point: A point that has at least \texttt{minPts} points (including itself) within its $\epsilon$-neighbourhood. Core points reside in high-density regions and form the structural backbone of clusters.}

\item{Border point: A point that falls within the $\epsilon$-neighbourhood of a core point but does not itself meet the \texttt{minPts} threshold to be considered a core point. These points lie at the periphery of dense regions.}

\item{Noise point (or outlier): A point that is neither a core point nor a border point. Such points do not fall within the $\epsilon$-neighbourhood of any core point and lies in low-density, unclustered regions.}
\end{itemize}

Owing to DBSCAN's sensitivity to variations in local density, its applicability can be limited in real-world scenarios where data are often heterogeneous and contaminated by noise. To address these challenges, Hierarchical Density-Based Spatial Clustering of Applications with Noise (HDBSCAN) extends DBSCAN by integrating hierarchical density estimation with a robust cluster extraction framework \cite{hdbscan1,hdbscan2,McInnes2017}. Unlike DBSCAN, which relies on a fixed global density threshold, HDBSCAN constructs a condensed cluster tree from the mutual reachability graph of the data, capturing a nested hierarchy of clusters across varying density levels. This multi-scale approach enables the identification of meaningful structures in data with non-uniform density. In light of these advantages, we adopt HDBSCAN to analyze the population structure of FRBs. To determine the optimal HDBSCAN hyperparameters, we employ a grid search approach to maximize classification performance. The hyperparameter settings used in our study are summarized in Table \ref{tab:HDBSCAN}.

\begin{table}[htbp]
\centering
\caption{List of HDBSCAN hyperparameters.}\label{tab:HDBSCAN}
\arrayrulewidth=1.0pt
\renewcommand{\arraystretch}{1.3}
{\begin{tabular}{cc} 
\hline\hline 
Name & Value\\\hline
min$\_$cluster$\_$size & 150\\
min$\_$samples & 2 \\
cluster$\_$selection$\_$epsilon & 0 \\
max$\_$cluster$\_$size & 0 \\
metric & `\ euclidean' \\
alpha & 1 \\
p & None \\
algorithm & `\ best' \\
leaf$\_$size & 40 \\
memory & Memory(cachedir = None,verbose = 0) \\
approx$\_$min$\_$span$\_$tree & True \\
gen$\_$min$\_$span$\_$tree & False \\
core$\_$dist$\_$n$\_$jobs & 4 \\
cluster$\_$selection$\_$method & `\ eom' \\
allow$\_$single$\_$cluster & False \\
prediction$\_$data & False \\
match$\_$reference$\_$implementation & False \\
\hline
\end{tabular}}
\end{table}

\section{Results and analysis}\label{sec:results}

\subsection{Results of the full features}\label{subsec: results_full}

\begin{figure}[ht!]
\centering
\includegraphics[width=0.45\textwidth]{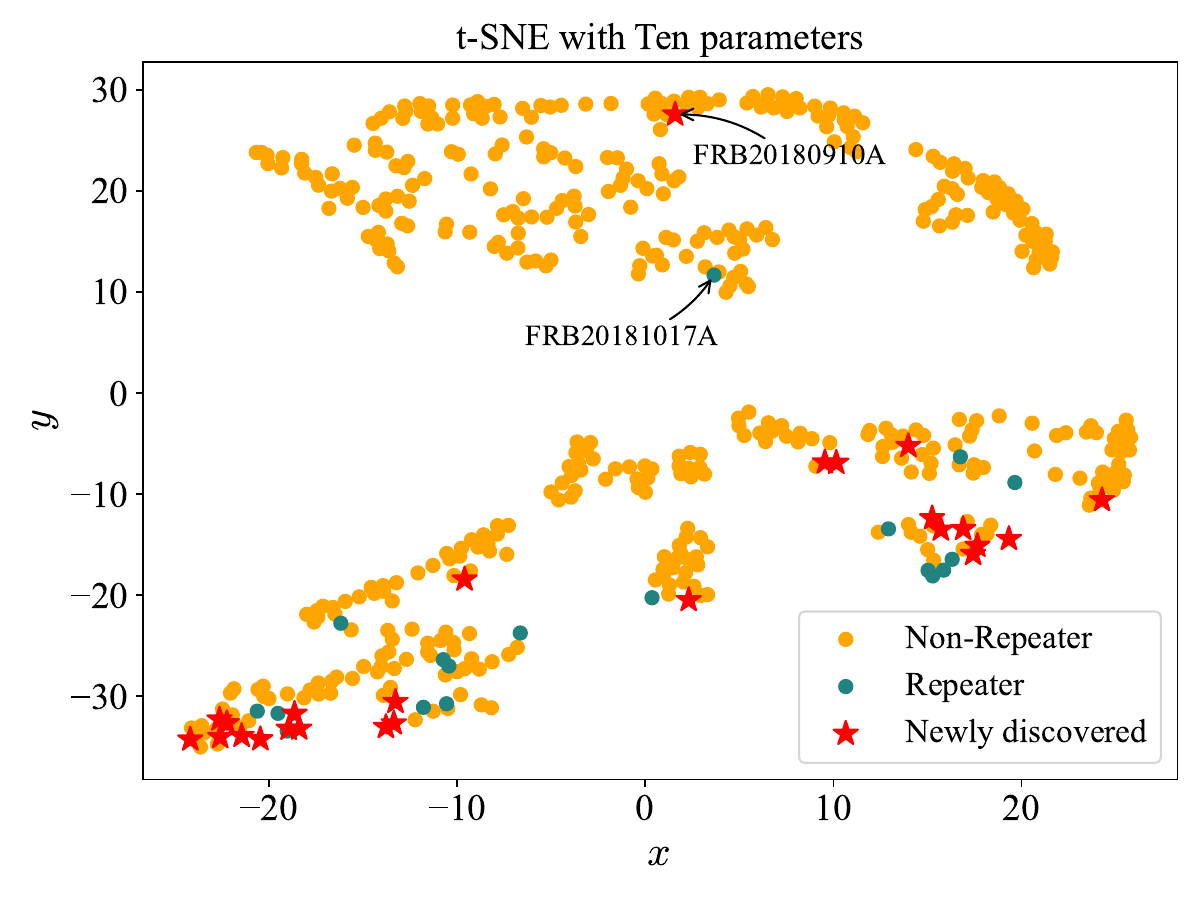}
\includegraphics[width=0.45\textwidth]{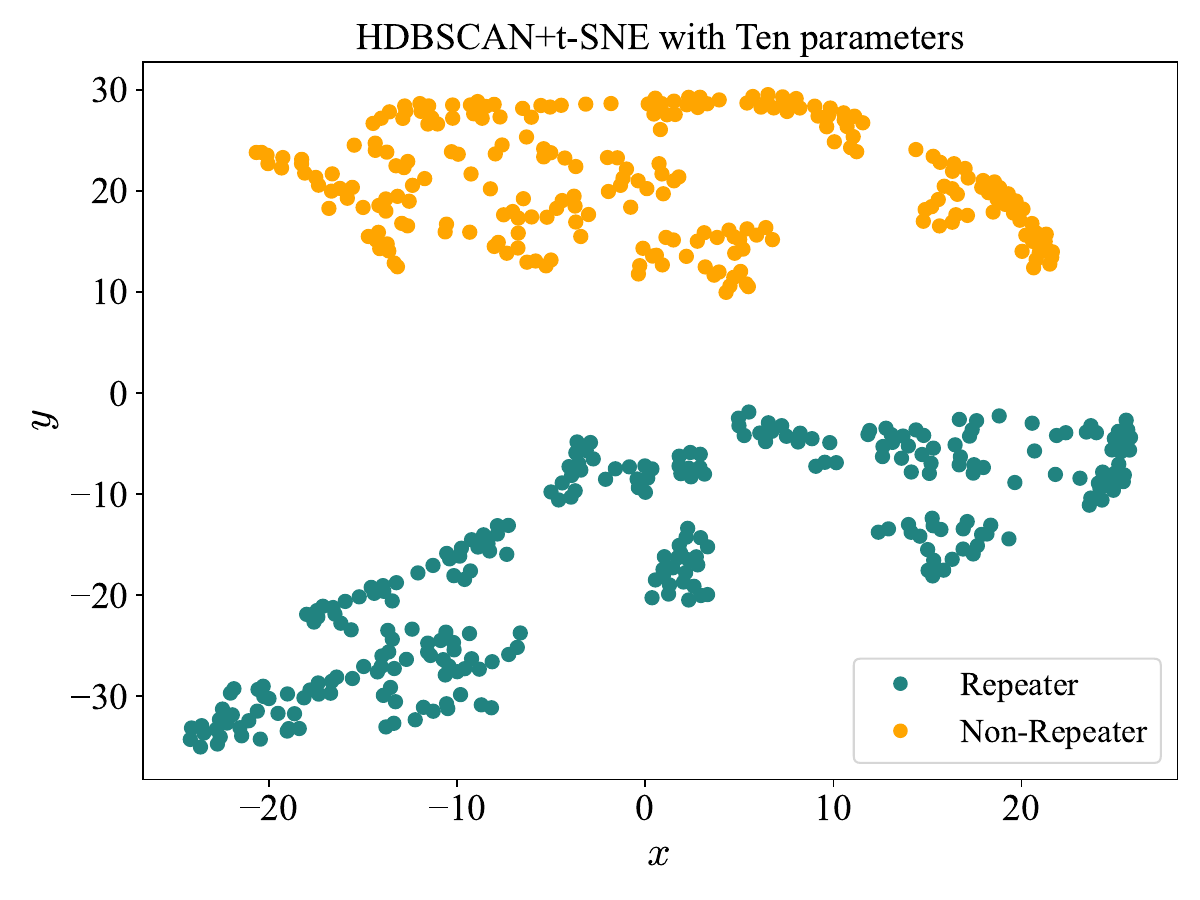}
\caption{Dimensionality reduction and clustering results with ten features of FRBs.}
\label{fig:result_tSNE_10}
\end{figure}

We first apply t-SNE to a ten-dimensional parameter space encompassing both observed and derived properties of repeating and non-repeating sources. The resulting two-dimensional projection is shown in the left panel of Fig.\ref{fig:result_tSNE_10}, where non-repeaters (orange) and repeaters (green) occupy distinct regions of the embedded space. Notably, 25 newly identified repeaters from the CHIME/FRB 2023 repeater catalog (including 6 FRBs misclassified as non-repeaters in the first CHIME/FRB catalog) are highlighted as red stars. Most known repeaters cluster in the lower quadrant of the embedding, with the exception of FRB20181017A and FRB20180910A, while non-repeating bursts form several well-defined and spatially separated groupings. The presence of non-repeaters within or near repeaters suggests possible misclassification.

The HDBSCAN algorithm is applied to the two-dimensional embedding to examine the latent structure uncovered by the t-SNE projection. The clustering results, presented in the right panel of Fig.\ref{fig:result_tSNE_10}, reveal a robust bifurcation of the FRB sample into two principal groups. The first cluster, shown in orange, is dominated by non-repeating FRBs and is hereafter referred to as the ``non-repeater cluster". Notably, this group also includes the previously known repeater FRB20181017A and the recently confirmed repeater FRB20180910A, both of which are misclassified within the non-repeater population \---- an anomaly that will be addressed in detail in Section {\ref{subsec:results_intrinsic}}. The second cluster, shown in green, encompasses all remaining repeaters in the sample, and is accordingly designated as the ``repeater cluster". Intriguingly, five out of six newly identified repeating sources (except for FRB20180910A), which are originally labeled as non-repeaters in the CHIME/FRB catalog, coincide spatially with this repeater cluster. This spatial congruence provides compelling evidence that certain sources previously considered non-repeating may, in fact, possess intrinsic properties akin to known repeaters. The ``repeater cluster" contains $41$ confirmed repeaters and $230$ candidate repeaters, which implies that more than half FRB sources may be intrinsic repeaters. These results support the hypothesis that repetition may be a function of observational limitations rather than a fundamental dichotomy in progenitor types. As such, non-repeating FRBs situated within the repeater cluster can be treated as repeater candidates, warranting further targeted follow-up.

\begin{figure}[htbp]
\centering
\includegraphics[width=0.5\textwidth]{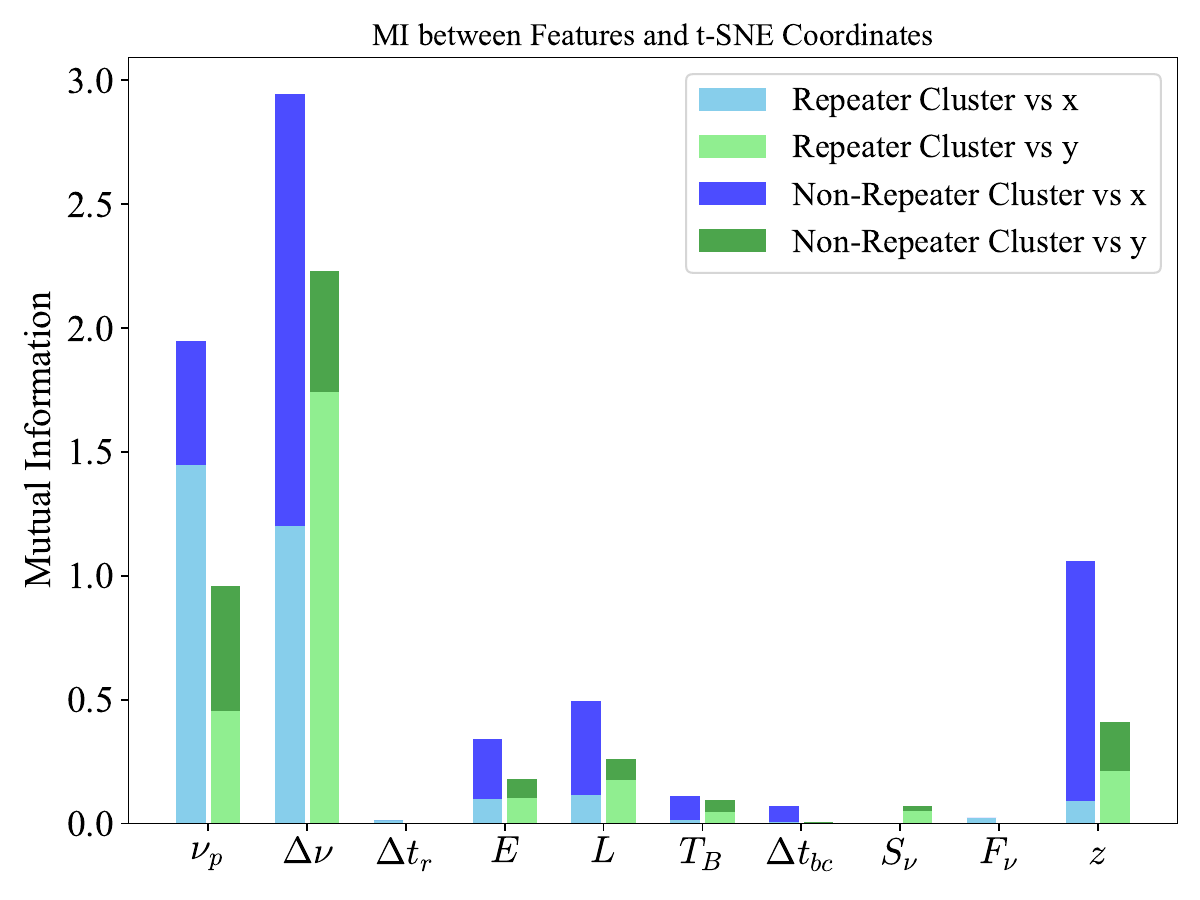}
\caption{Feature correlation of t-SNE + HDBSCAN with ten features.}
\label{fig:MI_10}
\end{figure}

To identify the key physical parameters underpinning the unsupervised clustering of FRBs, we quantify the dependency between each of the ten features and the cluster structure derived from the t-SNE embedding followed by HDBSCAN segmentation. This is achieved through the computation of mutual information (MI) \cite{Battiti298224,ROVIRA2022135250}, a non-negative measure of statistical dependence between two variables. By definition, MI vanishes when variables are independent and increases with stronger dependency, making it a robust and model-free metric for feature relevance. We employ the MI regression implementation in the scikit-learn library to estimate the mutual information between each feature and the two-dimensional t-SNE coordinates. For each identified cluster, MI scores are computed separately with respect to the two projection axes. The resulting MI distributions are shown in Fig.\ref{fig:MI_10}, where each feature is represented by a pair of bars corresponding to its relevance along the $x$- and $y$-axes of the embedding. For the non-repeater cluster, peak frequency, rest-frame frequency width, and redshift emerge as the most informative features. In contrast, the repeater cluster exhibits comparatively lower dependence on redshift. Taken together, the MI analysis indicates that peak frequency and rest-frame frequency width are the two most influential features governing the overall geometry of the low-dimensional representation.

\subsection{Results of the intrinsic features}\label{subsec:results_intrinsic}

\begin{figure}[ht!]
\centering
\includegraphics[width=0.45\textwidth]{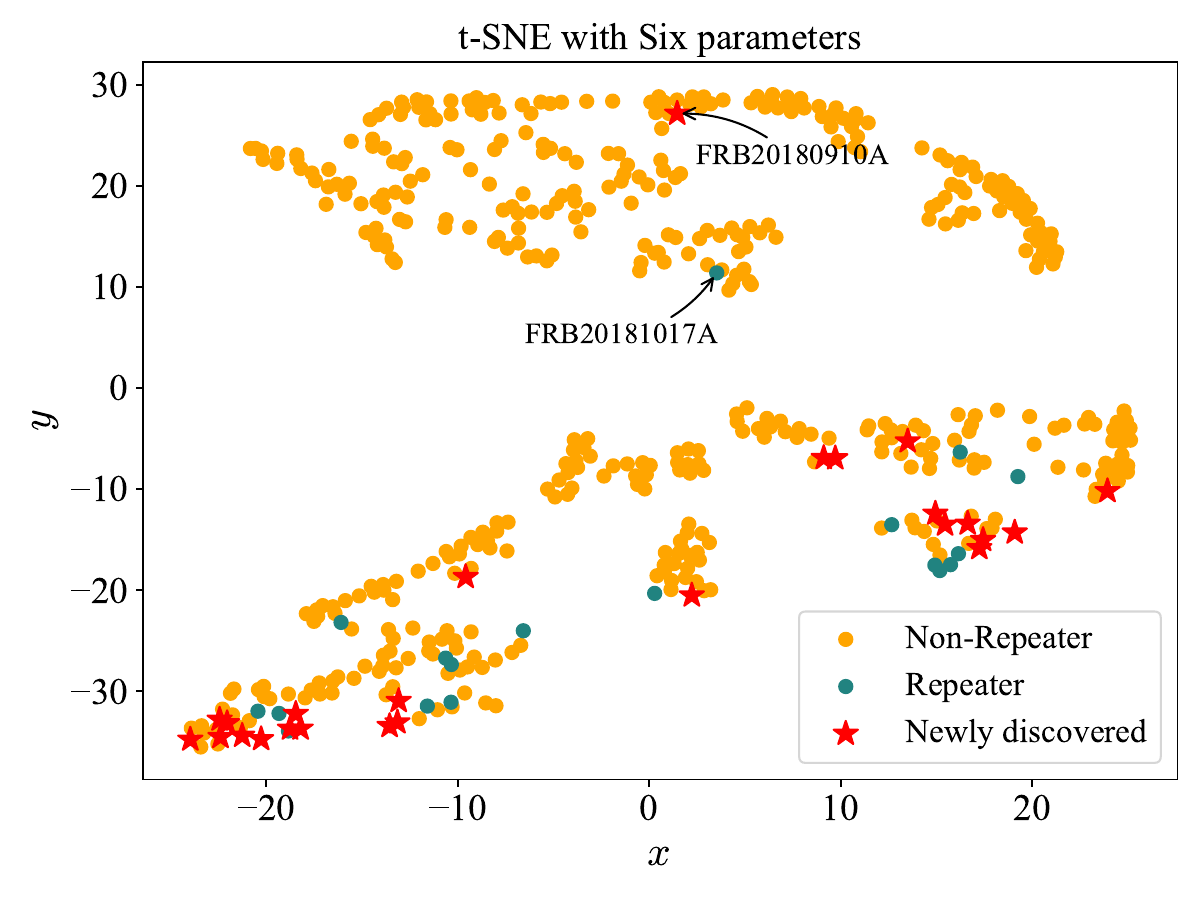}
\includegraphics[width=0.45\textwidth]{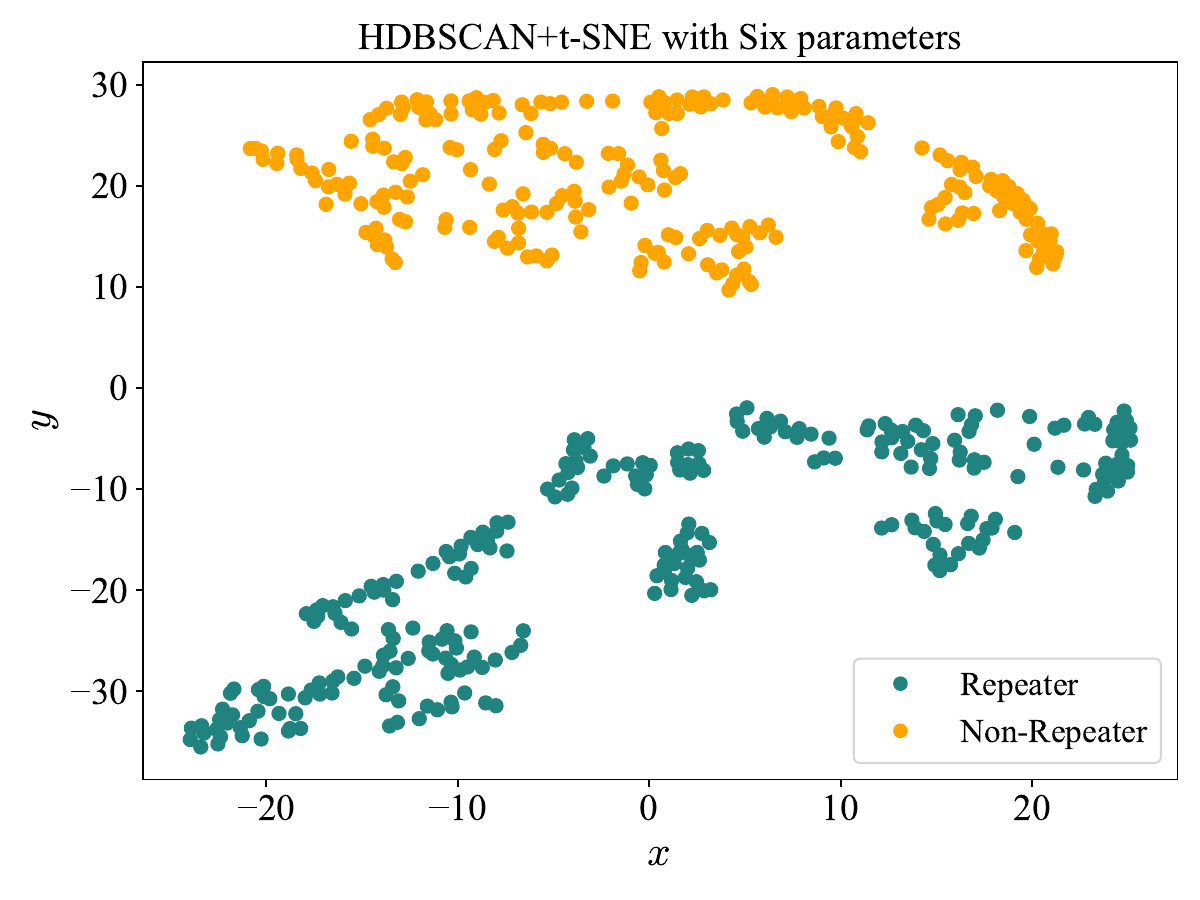}
\caption{Dimensionality reduction and clustering results with six features of FRBs.}
\label{fig:result_tSNE_6}
\end{figure}

The MI scores calculated in the ten-dimensional parameter space demonstrate that not all features exhibit significant contributions to the clustering structure of FRB populations. Notably, some features are non-intrinsic properties: redshift is a manifestation of the distance of FRB source, flux and fluence are inversely proportional to the square of distance, and boxcar width is an instrumental representation of burst duration that correlates with rest-frame temporal width. To optimize the feature space, we exclude these four non-intrinsic parameters exhibiting consistently low MI scores, thereby retaining six intrinsic features: peak frequency, rest-frame frequency width, rest-frame temporal width, burst energy, luminosity, and brightness temperature. This refined feature set is subsequently employed to recalculate the t-SNE dimensionality reduction and implement HDBSCAN clustering analysis, with all hyperparameters keep the same as in the ten-feature case. The updated results based on the six intrinsic features are shown in Fig.\ref{fig:result_tSNE_6}.

\begin{figure}[ht!]
\centering
\includegraphics[width=0.5\textwidth]{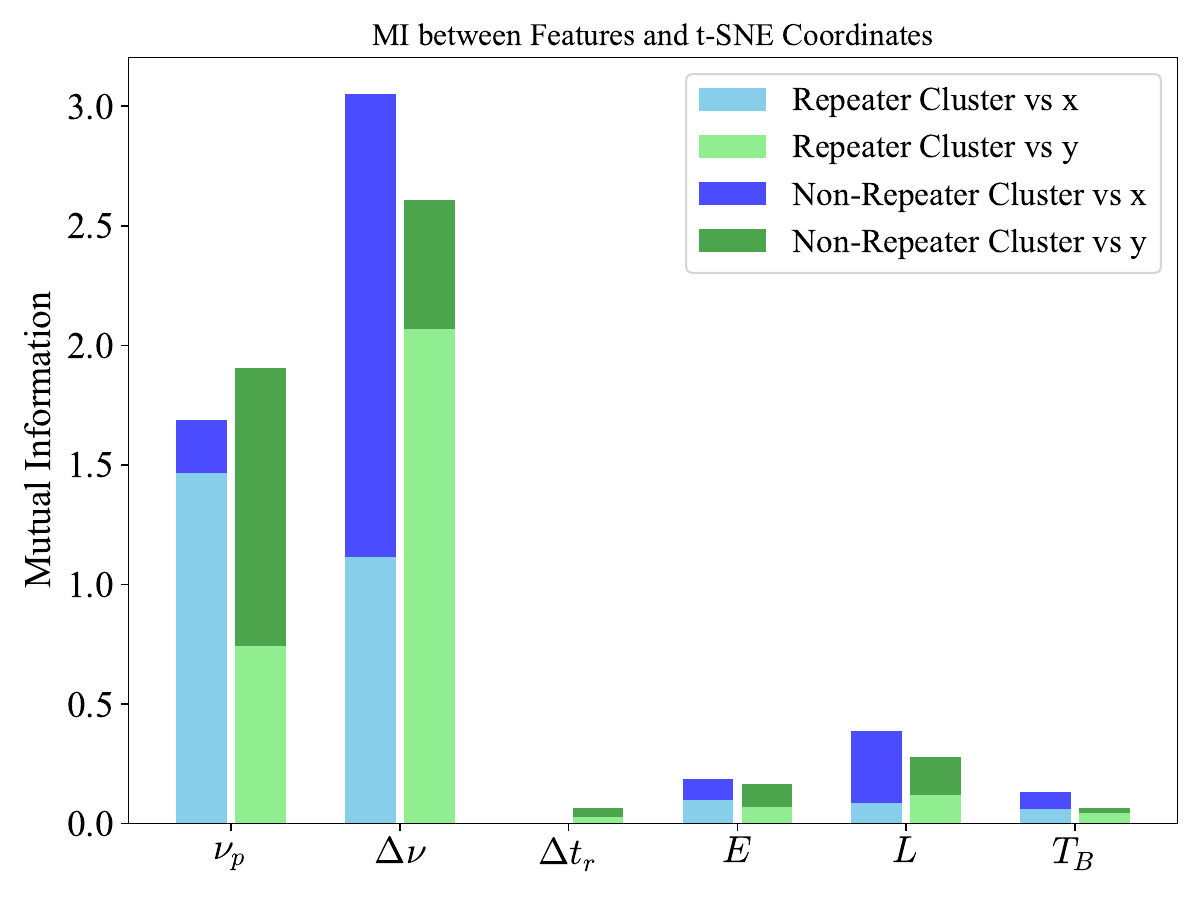}
\caption{Feature correlation of t-SNE + HDBSCAN with six features.}
\label{fig:MI_6}
\end{figure}

The two-dimensional projection of the refined six-dimensional parameter space preserves the overall topological separation identified in the original analysis. Repeaters (green) remain predominantly confined to the lower region of the embedding, whereas non-repeaters (orange) continue to cluster in the upper. The newly confirmed repeater FRB 20180910A and the previously known repeater FRB 20181017A are again located within the non-repeater-dominated region, suggesting these sources remain outliers irrespective of feature dimensionality. Application of HDBSCAN to the updated t-SNE embedding reproduces the previous segmentation, yielding two principal clusters with comparable spatial boundaries. This outcome confirms that the excluded features \---- redshift, flux, fluence, and boxcar width \---- play a minimal role in the latent separability of FRB populations. We further recompute the MI between feature values and t-SNE coordinates for both clusters in Fig.\ref{fig:MI_6}. It is found that the rest-frame frequency width and peak frequency continue to dominate, while the MI scores for the other features remain comparatively low.

\begin{figure}[ht!]
\centering
\includegraphics[width=0.45\textwidth]{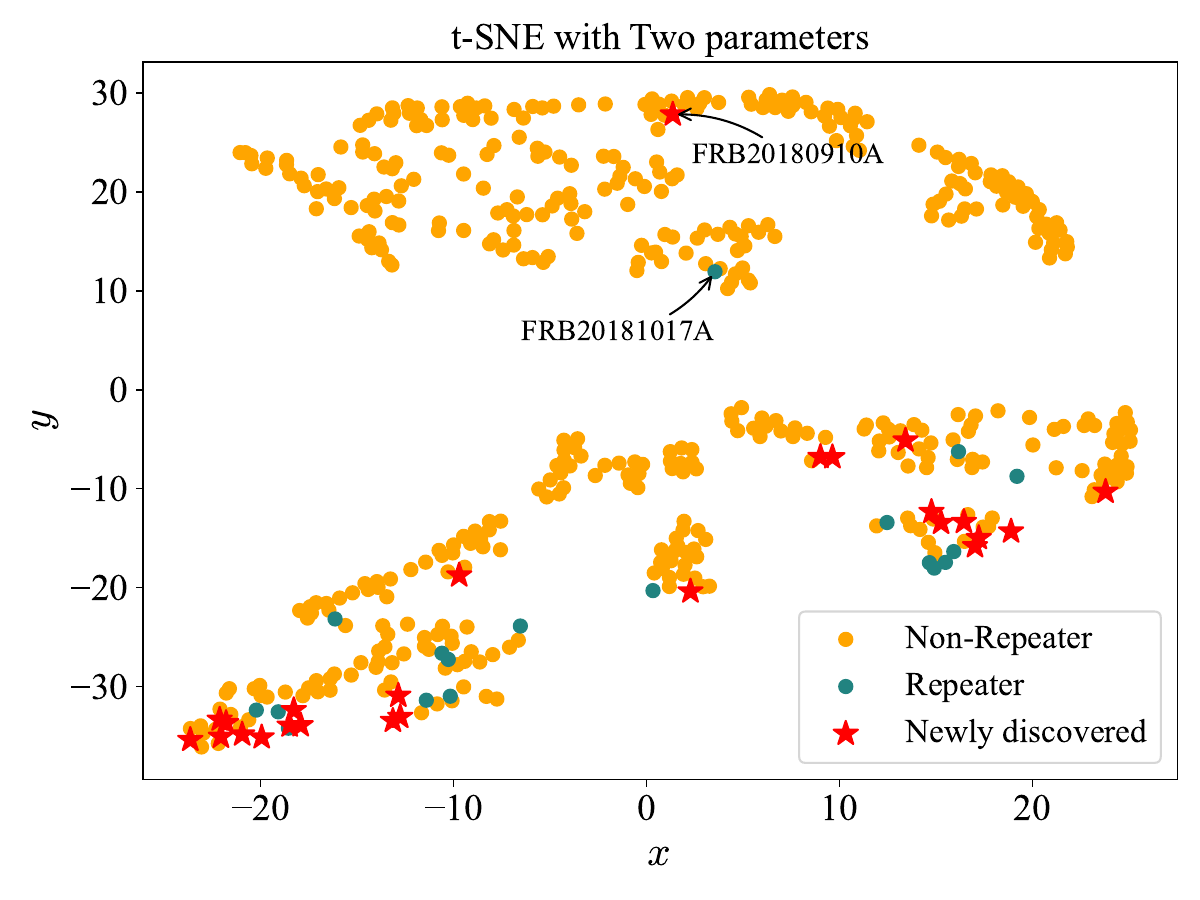}
\includegraphics[width=0.45\textwidth]{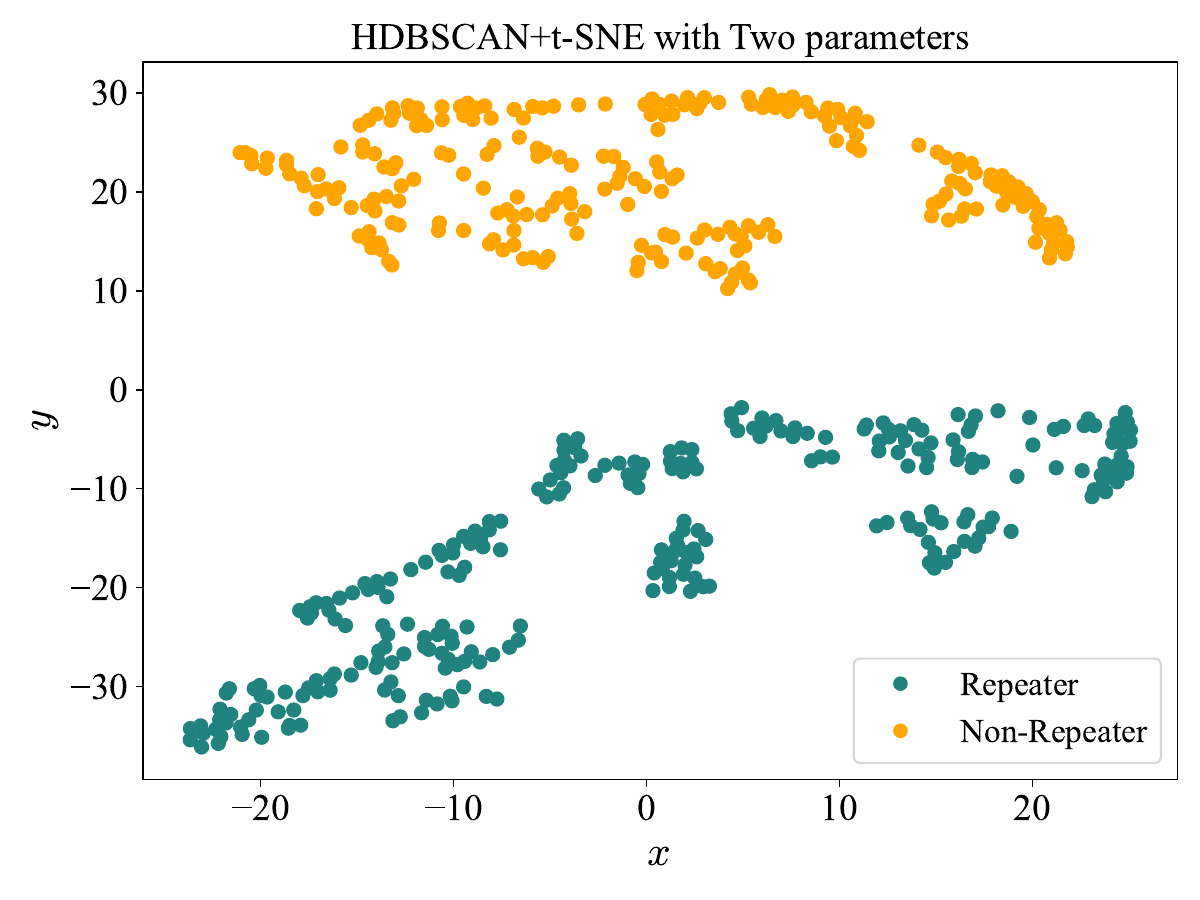}
\caption{Dimensionality reduction and clustering results with two features of FRBs.}
\label{fig:result_tSNE_2}
\end{figure}

Given the consistently highest MI scores for peak frequency and rest-frame frequency width, we restrict the input feature space to these two parameters. t-SNE and HDBSCAN are reapplied to this reduced set, with the resulting embeddings shown in Fig.\ref{fig:result_tSNE_2}. Remarkably, the fundamental separation between repeating and non-repeating FRBs is maintained: repeaters cluster in the lower quadrant, while non-repeaters dominate the upper. Despite the sharp dimensionality reduction, the core structural division observed in the ten- and six-feature analyses persists, suggesting robustness against feature pruning. HDBSCAN likewise continues to recover two principal populations, confirming the strong discriminatory power of peak frequency and rest-frame frequency width alone. Comparing Figures \ref{fig:result_tSNE_10}, \ref{fig:result_tSNE_6}, and \ref{fig:result_tSNE_2} reveals that classification performance remains largely unchanged even when the input feature dimensionality is sharply reduced. This consistency stems from the dominant role of two key features: $\nu_p$ and $\Delta\nu$, as evidenced by their exceptionally high MI scores in Figures \ref{fig:MI_10} and \ref{fig:MI_6}. Note that Figures \ref{fig:result_tSNE_10}, \ref{fig:result_tSNE_6}, and \ref{fig:result_tSNE_2} are not identical, though the differences are minimal. The slight shift of each data point across the three feature sets (10, 6, and 2) is visually indistinguishable because of the large scales of the x and y axes in the embedding plane.

\begin{figure}[ht!]
\centering
\includegraphics[width=0.5\textwidth]{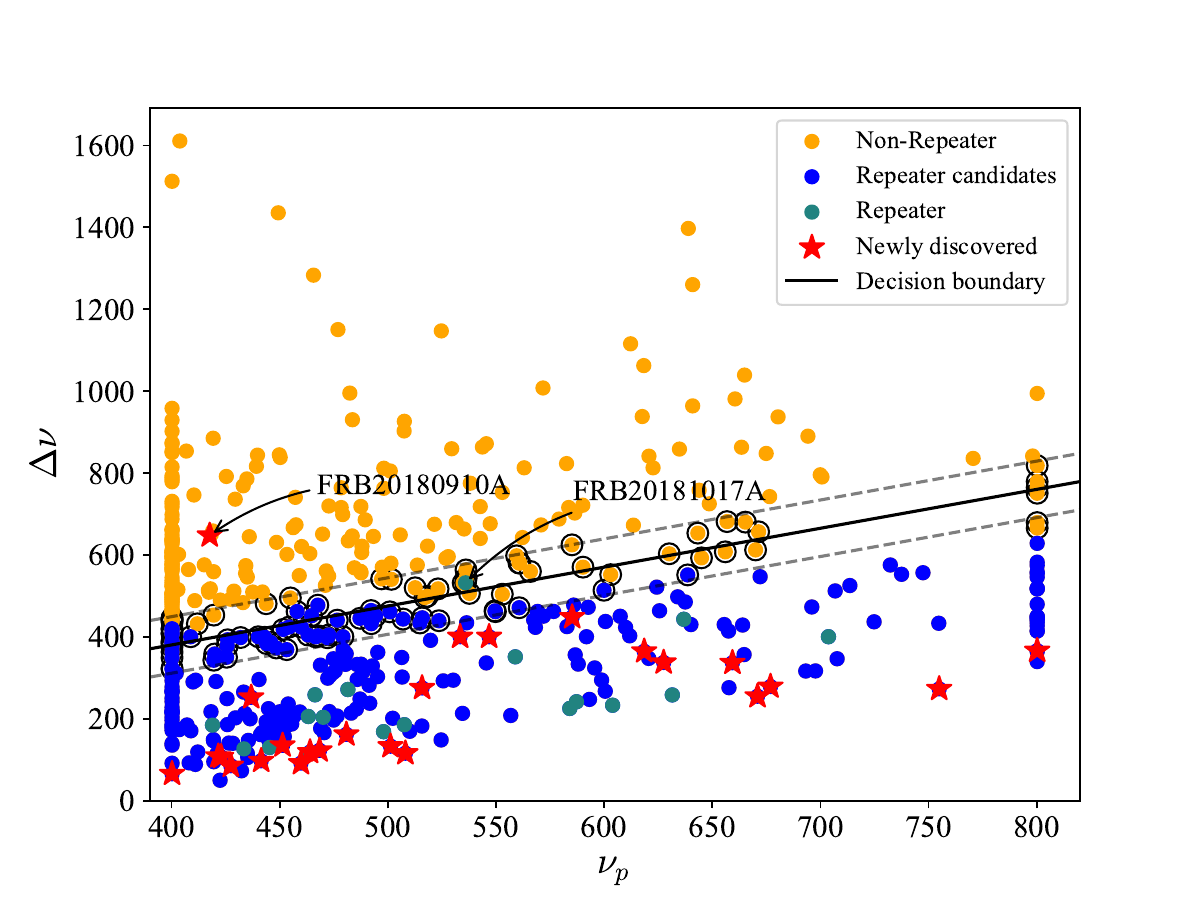}
\caption{The 2D plane of peak frequency $\nu_p$ and rest-frame frequency width $\Delta \nu$.}
\label{fig:f_Delta_nu}
\end{figure}

To facilitate direct interpretation of the classification boundary within the original feature space, we analyze the scatter plot of $\nu_p$ versus $\Delta \nu$ shown in Fig.\ref{fig:f_Delta_nu}, where a discernible separation between repeating and non-repeating FRBs emerges. The support vector machine (SVM) algorithm was implemented to construct the maximum-margin hyperplane that optimally separates the two classes through margin maximization. The resulting decision boundary, shown as a solid black line in Fig.\ref{fig:f_Delta_nu}, is given by $\Delta \nu = 0.95 \nu_p + 1.30$. Support vectors \---- samples located on the dashed margin lines \---- govern the position of this boundary. It is worth mentioning that, in the ideal linearly separable scenario, all training samples are expected to lie outside the margin boundaries, such that the perpendicular distance from each sample to the decision hyperplane is at least one. That is, points residing within the margin region violate the optimality constraints: although some may remain correctly classified, they lack sufficient separation from the decision boundary, while others are misclassified. For instance, the previously known repeater FRB20181017A is misclassified by the model as a non-repeater, so that lies within the margin. Surprisingly, the newly confirmed repeater FRB20180910A, which shows a high $\Delta \nu$, falls outside the expected region and is also misclassified as a non-repeater. Compared with higher-dimensional embeddings using six or ten features, this two-feature model provides a more interpretable, physically motivated classification boundary with no appreciable loss in performance. These findings highlight peak frequency and rest-frame frequency width as the two key parameters encoding the physical differences between FRB populations.

\subsection{Evaluating the model performance}

In machine learning, precision and recall are two key metrics commonly used to evaluate classifier performance. Precision measures the proportion of predicted positive instances that are correctly identified, corresponding to the accuracy of positive predictions. Recall, by contrast, quantifies the fraction of true positive instances correctly identified among all actual positives, and is also referred to as sensitivity or the true positive rate. In our work, the positive is repeaters, while the negative is non-repeaters. Given that the classification of non-repeating FRBs may be affected by limited follow-up observations and the incomplete detection of repetition, maximizing recall is particularly important to ensure that potential repeaters are not inadvertently excluded. Precision, in this context, would overly penalize the model for predicting repeaters among sources yet to be confirmed, thereby introducing bias against true but observationally incomplete repeaters. Recall thus offers a more meaningful measure of classifier effectiveness for FRB population studies.

The recall is defined as the ratio of true positive (TP), to the sum of true positive (TP) and false negative (FN),
\begin{equation}
    {\rm{Recall}}=\frac{\rm TP}{\rm TP+FN},
\end{equation}
where TP denotes the number of correctly classified repeaters, and FN represents repeaters incorrectly assigned to the non-repeater category. As the classification outcome in the two-feature space remains consistent with that derived from the higher-dimensional feature sets (ten or six features), we compute the recall based on the two-feature results. Our sample contains 43 repeaters (18 repeaters in the first CHIME/FRB catalog, and 25 newly discovered repeaters), two of which are misclassified as non-repeaters (FRB20180910A and FRB20181017A). The recall value is therefore $41/(41+2)=0.95$, indicating that even with only two input features, the model successfully recovers the majority of known repeaters. We confirmed that reducing the number of features from ten to two does not reduce recall. This performance underscores that the combination of peak frequency and rest-frame frequency width captures the essential physical distinctions between repeating and non-repeating FRBs.

\section{Discussion and conclusions}\label{sec:conclusions}

Despite the increasing detection rate of FRBs, their physical origins remain poorly understood. Current phenomenological classifications categorize FRBs into repeating and non-repeating populations based on burst recurrence. However, such distinctions are inherently limited by observational selection effects, notably instrumental sensitivity thresholds and finite monitoring durations, which may lead to misclassification. Addressing this challenge is crucial for advancing progenitor mechanism studies and developing accurate theoretical models. Traditional analytical methods struggle to handle the high-dimensional and heterogeneous nature of FRB datasets, prompting recent adoption of machine learning techniques, particularly unsupervised approaches, to uncover latent patterns in these complex data. A critical limitation of prior studies lies in their treatment of individual bursts (whether from repeaters or non-repeaters) as independent events, an approach that introduces data redundancy and risks biasing classification outcomes.

In this work, we reanalyzed the first CHIME/FRB catalog and the CHIME/FRB 2023 repeater catalog through an unsupervised machine learning framework optimized for capturing intrinsic FRB population characteristics. To mitigate bias from intrinsic burst correlations within sources, only one burst from each source was utilized. First, the t-SNE algorithm was employed to reduce the dimensionality of the FRB parameter space onto a two-dimensional manifold. Subsequently, HDBSCAN was applied to the low-dimensional embedding to identify natural groupings without a prior assumptions regarding the number of clusters. One cluster encompasses all known repeaters except for FRB20181017A and FRB20180910A, and is henceforth referred to as the ``repeater cluster". The second cluster is predominantly composed of non-repeating sources. Notably, five out of six FRBs initially classified as non-repeating sources, but recently confirmed as repeaters, are naturally embedded within the repeater cluster. A notable exception is FRB20180910A, which our analysis categorizes as a non-repeater despite its recent confirmation as a repeating source. This source is particularly distinctive because its ratio of $\Delta\nu$ to $\nu_p$ is significantly larger than that of other detected or predicted repeating FRBs. This outlier property separates FRB20180910A from the rest of the sample and may suggest a distinct emission mechanism or physical environment for this source. For the other outlier, FRB20181017A, its misclassification can be primarily attributed to its location near the SVM decision boundary in the $\Delta\nu$–$\nu_p$ parameter space, as illustrated in Fig.\ref{fig:f_Delta_nu}. Importantly, we demonstrated that the clustering structure remains stable even when the dimensionality of the feature space is reduced from ten to two, indicating the robustness of the identified separation.

MI analysis revealed that the peak frequency $\nu_p$ and rest-frame frequency width $\Delta \nu$ are the most informative parameters shaping the low-dimensional representation, in agreement with earlier findings \cite{Zhu-Ge:2022nkz}. Scatter plots in the $\Delta \nu$–$\nu_p$ plane reveal that the two clusters can be separated by a straight line $\Delta \nu = 0.95 \nu_p+1.30$, with repeaters tend to exhibit systematically narrower $\Delta \nu$ compared to non-repeaters. The outlier sources FRB20181017A and FRB20180910A are misclassified primarily due to their $\Delta \nu$ values exceeding the empirical decision boundary determined via a SVM model. Overall, our findings suggest that a simplified classification scheme based solely on $\Delta \nu$ and $\nu_p$ can effectively distinguish between repeating and non-repeating FRBs. This approach offers a promising framework for the future analysis of expanding FRB catalogs.

Unsupervised machine learning has previously been employed to classify CHIME/FRBs, as demonstrated by Zhu-Ge et al. \cite{Zhu-Ge:2022nkz} and Qiang et al. \cite{Qiang:2024lhu}.  Although the method used in our work bears similarity to these prior studies, it incorporates several novelty: First, our dataset selection criteria differ. Previous analyses typically utilized the full CHIME/FRB catalog, including all bursts from repeating sources and all sub-bursts from apparently non-repeating sources. By contrast, our study employs only one burst per source. Second, our feature selection approach diverges. While we initially adopt the same ten features used in earlier work, we demonstrate that the number of features can be reduced to two while maintaining virtually unchanged model performance. This simplification not only streamlines classification but also reveals the key physical properties governing FRB categorization. Third, our results exhibit significant differences. Unlike prior analyses that partitioned FRBs into multiple categories, we show that with appropriate hyperparameter tuning, our method robustly segregates the sample into two distinct clusters. Finally, we report a critical new finding: repeaters and non-repeaters exhibit linear separability in the $\Delta\nu$–$\nu_p$ plane. This provides a method for FRB classification without dimensionality reduction, suggesting a fundamental distinction between the two populations.

\vspace{5mm}
\centerline{\rule{80mm}{0.5pt}}
\vspace{2mm}

\bibliographystyle{cpc}
\bibliography{reference}

\end{document}